Su dongcai

# Compressed sensing with partially corrupted Fourier measurements

Dongcai Su at 2016-6-7

## Abstract

This paper studies a data recovery problem in compressed sensing (CS), given a measurement vector $b$ with corruptions: $b = Ax^{(0)} + f^{(0)}$, can we recover $x^{(0)}$ and $f^{(0)}$ via the reweighted $\ell_1$ minimization: $\min_{x,f} \|x\|_1 + \lambda \|f\|_1, s.t. Ax + f = b$ ? Where the $m \times n$ measurement matrix A is a partial Fourier matrix, $x^{(0)}$ denotes the $n$ dimensional ground true signal vector, $f^{(0)}$ denotes the $m$-dimensional corrupted noise vector, where a positive fraction of entries in the measurement vector $b$ are corrupted by the non-zero entries of $f^{(0)}$. This problem had been studied in literatures [1-3], unfortunately, certain random assumptions (which are often hard to meet in practice) are required for the signal $x^{(0)}$ in these papers. In this paper, we show that $x^{(0)}$ and $f^{(0)}$ can be recovered exactly by the solution of the above reweighted $\ell_1$ minimization with high probability provided that $m \geq O\left(\left\|x^{(0)}\right\|_0 \log^2(n)\right)$[1] and $n$ is prime, here $\left\|x^{(0)}\right\|_0$ denotes the cardinality (number of non-zero entries) of $x^{(0)}$. Except the sparsity, no extra assumption is needed for $x^{(0)}$.

## 0. Introduction

In a traditional compressed sensing framework, a n-dimensional signal $x^{(0)}$ is recovered from its $m$ linear projection measurement $b$, where $b = Ax^{(0)}$. Here, the $m$ dimensional vector $b$ is called the measurement vector, the $m \times n$ matrix $A$ is called the measurement matrix (or sensing matrix). When the ground true signal is sparse or approximately sparse, and the measurement matrix satisfies some properties, e.g., the restricted isometry property (RIP), then $x^{(0)}$ is proven to be recovered exactly from the below $\ell_1$ minimization:

$$\min_x \|x\|_1, s.t. Ax = b \tag{0.1}$$

---

[1] This result can be further improved to be $m \geq O\left(\left\|x^{(0)}\right\|_0 \log(n)\right)$ by applying of weak RIP in [4] E. J. Candes and Y. Plan, "A probabilistic and RIPless theory of compressed sensing," *Information Theory, IEEE Transactions on,* vol. 57, pp. 7235-7254, 2011.

Some typical classes of measurement matrix which satisfy RIP are, namely, sub-Gaussian matrix [5], and partial Bounded Orthogonal System (BOS) [6] where the matrix is formed by selecting rows uniformly at random from a BOS matrix.

Typical recovery guarantee of (0.1) with these kinds of measurement matrices are listed as following : 1) when A is sub-Gaussian matrix, then the solution of (0.1) exactly recover $x^{(0)}$ with high probability, provided that $m \geq O\left(k \ln\left(\frac{en}{k}\right)\right)$ [5]. 2) when A is partial BOS, then the recovery of (0.1) is exact with high probability given that $m \geq O\left(k \ln^4(n)\right)$ [7]. Here, $k$ is a positive constant that stands for the upper bound of the cardinality of $x^{(0)}$, e.g., $k = \max\left\{1, \|x^{(0)}\|_0\right\}$.

Although the recovery guarantee of (0.1) is promising, when the measurement vector $b$ involved some corruptions or irrelevant measurements in its entries (which unfortunately is prevalent in practice [2, 3, 8]), exact recovery of (0.1) fails to be guaranteed.

Motivated by this, another line of works (called compressed sensing with corruptions [3] or corrupted sensing [9]), try to model the corruptions as a sparse $m$ dimensional vector $f^{(0)}$, and propose to recover $x^{(0)}$ and $f^{(0)}$ through the below reweighted $\ell_1$ minimization:

$$\min_{x,f} \|x\|_1 + \lambda \|f\|_1, s.t. Ax + f = b \tag{0.2}$$

Where $\lambda > 0$ in (0.2) serves as a weighting factor, in the case when A is sub-Gaussian matrix, the recovery guarantee of (0.2) is almost as promising as those of (0.1). E.g., [10] show that the recovery of (0.2) is possible when a constant fraction of the measurement vector $b$ is corrupted by $f^{(0)}$, provided that $m \geq O\left(\|x^{(0)}\|_0 \log(n)\right)$. This is not surprising, since $[A, I]$ satisfies the RIP with high probability [11].

Does this hold when A is partial BOS matrix? More specifically, we'd like to ask, under what circumstance, the recovery guarantee of (0.2) is as good as those of (0.1) when A is partial BOS matrix?

At a first glance, $\wedge(s_f^c)$ being a random subset of $[n]$ seems to be a necessary requirement, here $\wedge$ denotes the row indices set (associate to the Fourier basis matrix) of A, $s_f$ denotes the support of $f^{(0)}$, with $s_f^c$ to be the complement of $s_f$. Since otherwise, successful recovery by (0.2) implies the successful recovery by (0.1) with $\wedge$ unnecessary to be a random subset[II] of $[n]$, which leads to a stronger conclusion than typical CS literatures [1, 7]. So a more practical question we attempt to seek answer to in this paper is presented below:

**Q(0.1)** *If A is a partial BOS matrix, whose rows are chosen randomly and uniformly from a BOS, assuming $\wedge(s_f)$ is a random subset of $[n]$, and $\dfrac{|s_f|}{m} = \gamma_c \in [0,1)$, can (0.2) recover $x^{(0)}$ and $f^{(0)}$ exactly with high probability?*

We first argue that if we choose the weighting factor $\lambda \leq 1$ in (0.2) then the answer to Q (0.1) is negative.

To see this, consider a special instance where A is a Fourier matrix (which is a special case of partial BOS), let $x^{(0)} = d$ be a Dirac comb vector, and $f^{(0)} = 0$ be a zero vector.

Where the Dirac comb vector $d$ [12] is a $n$-dimensional vector defined as,

$$d(t) = \begin{cases} 1, t = \sqrt{n}, 2\sqrt{n}, \ldots, n \\ 0, otherwise \end{cases} \tag{0.3}$$

Remarkably, the Fourier transform of $d$ (denoted by $\check{d}$) is also itself: $d = \check{d}$.

In this case, obviously (0.2) fails to admit $x^{(0)} = d, f^{(0)} = 0$ as its unique optimal solution, since $x = 0, f = d$ is a feasible point of (0.2) with smaller or equal objective value than $x^{(0)}, f^{(0)}$ when we choose $\lambda \leq 1$. Whether answers to Q (0.1) with $\lambda > 1$ exist remains to be

---

[II] See definition 1.1 for the definition of random subset.

an open problem to the best of our knowledge.

Having this simple counter example in mind, it is quite understandable that existing literatures study answers to problems like Q (0.1) require extra assumptions, e.g., [2, 3, 12] assume that the signs or supports of non-zero entries in $x^{(0)}$ to be independently random.

It's more practical appealing to allow the support and signs of $x^{(0)}$ to be arbitrary, in this paper, we proved that exact recovery of $x^{(0)}$ and $f^{(0)}$ by (0.2) is possible, provided that the assumption in Q (0.1) holds and $m \geq O\left(|s_x|\log^2(n)\right)$. Here, $s_x$ denotes the support of $x^{(0)}$, and $|s_x|$ denotes the cardinality of $s_x$. The only extra assumption we impose on Q (0.1) is that $n$ (which denotes the dimension of $x^{(0)}$) is prime, which excludes the existence of the Dirac comb vector in this case. We state our finding of this paper formally in theorem 1.1 of section 1.

Recently, [1-3] also study the problem of data recovery from corrupted Fourier measurements, which is similar to Q (0.1). We briefly summarize the difference between the results of the existing papers and the result we obtained in this paper as following, Candes [1] shows that when $A$ is full Fourier basis, choosing $\lambda = 1$ in (0.2), then exact recovery of $x^{(0)}$ and $f^{(0)}$ is possible when and both $\|x^{(0)}\|_0$ and $\|f^{(0)}\|_0$ is smaller than $O\left(n/\sqrt{\ln(n)}\right)$. More recently, Nguyen [2] and Li [3] shows that when $A$ is a partial BOS (where partial Fourier matrix can be treated as a special case), after choosing $\lambda < 1$ appropriately in (0.2), then recovery of $x^{(0)}$ and $f^{(0)}$ is possible even when a constant fraction of $b$ is corrupted by $f^{(0)}$, provided that $m \geq O\left(\|x^{(0)}\|_0 \ln^2(n)\right)$. Despite the promising theoretical results of these literatures, they all impose certain random assumption on $x^{(0)}$ and $f^{(0)}$, e.g., [1] assumes that the supports of $x^{(0)}$ and $f^{(0)}$ are uniformly random, [3] and [2] assume the signs of $x^{(0)}$ are random. On contrary, in this paper (see theorem 1.1), it require no assumption on $x^{(0)}$ except the sparsity, which renders our result to be more applicable in real applications.

Furthermore, our proof techniques are different from the previous literatures, e.g., in [2-4], they all prove their results by explicitly constructing a dual vector which satisfies the so called dual certification [13]. In this paper, firstly, it construct an "approximately feasible" dual vector $q_0$ through a golfing scheme presented section 2.4, then the vector $q_0$ is further modified to be a viable dual vector (which satisfies the dual certification) by adding a modification vector $\Delta q$, and the existence of this modification vector $\Delta q$ is argued by contradiction as stated in section 2.5. Therefore, the proof of the existence of the viable dual vector in this paper is inherently non-constructive, which is different from other approaches in literatures [3, 4], where the viable $q$ is directly constructed through golfing schemes.

We believe that the methods and ideas involved in the proof of this paper have importance and interest in their own right, and hopefully, they can give some insights of proving other problems in the CS community. This is the motivation of developing this paper.

The remaining parts of this paper are organized as following, in section 1 we proposed our main theoretical result, which is stated formally in theorem 1.1. Section 2 is dedicated to illustrate the proof of theorem 1.1: In section 2.1, we adopt a de-randomized technique [12, 14], which allows us to impose an extra assumption on the signs of $f^{(0)}$ without affecting the recovery guarantee of (0.2), section 2.2 states the dual certification on which we develop the proof of theorem 1.1. Next in section 2.3 it sketches the proof roadmap of the existence of a dual vector $q$ that satisfies the dual certifications as stated in section 2.2, with the "approximate feasible" dual vector $q_0$ constructed in section 2.4, and the existence of $\Delta q$ proved in section 2.5, respectively.   For convenience, we put all supporting and intermediate lemmas into Appendix A ~ Appendix D.

For the convenience of the readers, we introduce some notations that will be used in the remaining parts of this paper.

*Notations.* In this paper, $[n]$ denotes the indices set $\{1, 2, \ldots, n\}$ if $n$ represents an integer, $k + [n]$ denotes the translated indices set $\{k+1, k+2, \ldots, k+n\}$ if both $k$ and $n$ represent integers. $\|x\|_0$ denotes the number of non-zero entries of vector $x$ (or the cardinality of $x$) if $x$ denotes a vector. $|s|$ denotes the cardinality of set $s$ if $s$ denotes a set. $M^*$ denotes the conjugate of $M$ if $M$ denotes a matrix with complex entries. $I_N$ denotes a $N \times N$ identity matrix, where N is a positive integer.

# 1. Main results

Suppose we have a m-dimensional measurement vector $b$ which can be written as:

$$b = \tilde{A}x^{(0)} + f^{(0)} \tag{1.1}$$

Where $x^{(0)}$, $f^{(0)}$ denote the n-dimensional ground-true signal and m-dimensional corrupted noise, respectively. $\tilde{A}$ is a $m \times n$ column normalized partial Fourier basis:

$\tilde{A} = \sqrt{\frac{n}{m}} F(\wedge, [n])$, here the indices set $\wedge$ is defined as the random subset of [n] (see below definition 1.1 for the definition of random subset).

**Definition 1.1** *(random subset) An indices set $S = \{s_1, \ldots, s_m\}$ with cardinality m is called a random subset of [n], if $S \subseteq [n]$ and S is selected randomly and uniformly from all subset of [n] with cardinality m.*

In this paper, our goal is to recover $x^{(0)}$, $f^{(0)}$ through the below $\ell_1$ minimization:

$$\text{minimize}_{x,f} \|x\|_1 + \|f\|_1 \text{ s.t. } \lambda \tilde{A}x + f = b \tag{1.2}$$

E.g., let $\hat{x}, \hat{f}$ denote the solution of (1.2), then $\lambda \hat{x}$, $\hat{f}$ are estimations of $x^{(0)}$, $f^{(0)}$, respectively, here $\lambda > 0$ severs as a parameter of (1.2).

We prove that the solution of (1.2) can recover $x^{(0)}$ and $f^{(0)}$ exactly with high probability provided that $\tilde{A}$, $x^{(0)}$ and $f^{(0)}$ satisfy certain conditions, which are stated formally in below theorem 1.1.

**Theorem 1.1** If n is prime, $\tilde{A} = \sqrt{\frac{n}{m}} F(\wedge, [n])$ is a $m \times n$ column normalized partial Fourier matrix with $\wedge$ and $\wedge(s_f^c)$ to be random subsets of $[n]$, $m \geq c_m \ln(4n) \ln(2\varepsilon^{-1}) \sqrt{|s_x|}$,

$m \geq 318 c_\lambda^2 \beta^2 \ln(2n\varepsilon^{-1})|s_x|$ , $\sqrt{\frac{|s_f^c|}{2}} - |s_x| > 4.2\sqrt{3}\beta\sqrt{|s_x|}\sqrt{\ln\left(\frac{2n}{\varepsilon}\right)}$ and

$$\ln(4n) \geq \frac{\sqrt{3}}{2}\sqrt{\frac{1}{1-\gamma_c}}.$$

Setting $\lambda = \frac{c_\lambda}{\sqrt{\ln(2n/\varepsilon)}}$ in (1.2), with positive constant $c_\lambda \leq \frac{\sqrt{2}}{16}$, then the solution of (1.2) recover $x^{(0)}$ and $f^{(0)}$ exactly with probability at least $1-7\varepsilon$.

Where in theorem 1.1, $\gamma_c \in [0,1)$ is a constant representing the upper-bound of corruption ratio $\frac{|s_x|}{m}$, $c_m = 824\frac{\alpha}{1-\gamma_c}$ is some constant, with constant $\alpha \in (4,6)$, $\beta = \frac{1}{10c_\lambda}\sqrt{\frac{3\alpha}{1-\gamma_c}}\sqrt{\frac{3}{2}\frac{9}{8}}$ is a positive constant. $\wedge$ denotes the row indices (associated to the Fourier basis matrix) set of $\tilde{A}$, $s_f^c$ denotes the complement of the support of $f^{(0)}$ and $s_x$ denotes the support of $x^{(0)}$, $\varepsilon \in (0,1/7)$ is a positive constant.

**Remark:**

- Treating $\varepsilon^{-1} = O(n)$ in theorem 1.1, it shows that (1.2) can recover the ground true signal $x^{(0)}$ even when a positive fraction $\gamma_c$ (which can be arbitrarily close to 1) of the measurements are corrupted by $f^{(0)}$, provided that $m \geq O(|s_x|\log^2(n))$. This asymptotical bound can further be improved as $m \geq O(|s_x|\log(n))$ by using the weak RIP results in [4]. Here we obtain a slightly worse upper bound on $m$ because for the ease of conveying the core idea of our proof, we adapt the golfing scheme as presented in [6] chp 12, instead the one in [4].

- Constants present in theorem 1.1 are chosen merely for the convenience and the clarity of the proof, which are not the optimal ones.

## 2. Proof roadmap of theorem 1.1

The structure of this section is organized as follows, firstly, in section 2.1 (which follows directly from a de-randomize technique introduced in [14]), we show that imposing an extra random assumption the signs of $f^{(0)}(s_f)$ would not affect the generality of theorem 1.1, and hence in the remaining parts of this section, we'll develop our proof of theorem 1.1 by adding this extra assumption.

Secondly, we develop our proof for theorem 1.1 based on the dual certification as introduced in section 2.2, more specifically, if we can prove the existence of a dual vector $q$ which satisfies the dual certificate conditions as elaborated in lemma 2.2.2, then one can prove theorem 1.1.

To this end, firstly, in section 2.4, we construct a vector $q_0$ which approximately satisfies the condition in lemma 2.2.2, the vector $q_0$ is constructed by a golfing scheme which is derived from the golfing scheme introduced in [6]. And then in section 2.5, we prove the existence of a vector $\Delta q$, such that $q_0 + \Delta q$ satisfies the conditions in lemma 2.2.2. Finally, the conclusion of theorem 1.1 is proved by putting the results of section 2.1~2.5 altogether.

## 2.1 Derandomizing the signs of $f^{(0)}$

According the assumptions in theorem 1.1, $\wedge(s_f^c)$ is a random subset of $[n]$, one can easily conclude that $\wedge(s_f)$ is also a random subset of $[n]$. In this section, by applying a recently developed derandomizing technique in [14], we show that without affecting the recovery guarantee of (1.2) in general, one can impose an extract assumption on the corrupted noise vector $f^{(0)}$ that all non-zero entries in $f^{(0)}$ take signs {+1, -1} independently and uniformly at random.

Hence for convenience and without loose of generality, in the remaining section 2.2~2.5 of this paper where we present the proof of theorem 1.1, we assume the non-zero entries of $f^{(0)}$ in (1.2) also take signs $\{+1, -1\}$ uniformly at random.

**Definition 2.1.1 (Trimmed version vector) [14]** *we say that $f'$ is a trimmed version of $f$ if*

$$supp(f') \subset supp(f) \quad \text{and} \quad f'(i) = f(i) \quad \text{whenever} \quad i \in supp(f').$$

The below theorem is obtained following arguments similar in [14].

**Theorem 2.1.1 (elimination theorem)** *Suppose the solution to* (1.2) *with ground true* $x^{(0)}, f^{(0)}$

*is unique and exact, consider the ground true signal and corrupted noise with $x^{(0)}, f^{(0)'}$, where $f^{(0)'}$ is a trimmed version of $f^{(0)}$, then the solution to (1.2) with ground true $x^{(0)}, f^{(0)'}$ is also unique and exact.*

**Proof:**

Let $(\hat{x}, \hat{f})$ be the solution of (1.2) with ground true $x^{(0)}, f^{(0)'}$, then we have,

$$\|\hat{x}\|_1 + \|\hat{f}\|_1 \leq \|x^{(0)}\|_1 + \|f^{(0)}(s')\|_1 \tag{2.1.1}$$

Here, we denote $s'$ as the support set of $f^{(0)'}$, accordingly $s'^c$ denotes its complement set, then (2.1.1) implies,

$$\|\hat{x}\|_1 + \|\hat{f}\|_1 + \|f^{(0)}(s'^c)\|_1 \leq \|x^{(0)}\|_1 + \|f^{(0)}\|_1 \tag{2.1.2}$$

Applying the triangle rule, (2.1.2) yields,

$$\|\hat{x}\|_1 + \|\hat{f} + f^{(0)}(s'^c)\|_1 \leq \|x^{(0)}\|_1 + \|f^{(0)}\|_1 \tag{2.1.3}$$

By the optimality and unicity of $x^{(0)}, f^{(0)}$, one inevitably have $\hat{x} = x^{(0)}$ and $\hat{f} = f^{(0)'}$ according to (2.1.3), which proves the conclusion of the theorem.■

In theorem 1.1, it requires that $\wedge(s_f)$ is a random subset of $[n]$, where $\wedge$ denotes the rows indices set of A with respective to F (the $n \times n$ Fourier basis matrix). Since it's more convenient to work with the Bernoulli model in this section without affecting the recovery guarantee of theorem 1.1 [6]. We assume that elements in $\wedge(s_f)$ are sampled in $[n]$ according to a Bernoulli model with parameter $\rho \in (0,1)$ as described below.

***Definition 2.1.2 the Bernoulli model [14].*** *We say an indices set $\wedge$ is sampled according to a Bernoulli model with parameter $\rho \in (0,1)$ if it is constructed as following:*

*$j \in \wedge$ if and only if $\delta_j = 1$, where $\delta_j$ $j = 1, 2, \ldots, n$ are i.i.d Bernoulli random variables taking value one with probability $\rho$ and taking value 0 with probability $1 - \rho$.*

***Theorem 2.1.2*** *Suppose $x^{(0)}$ obeys the condition of theorem 1.1 and that elements in $\wedge(s_f)$ are chosen according to a Bernoulli model with parameter $2\rho$, and the signs of entries in $f^{(0)}$* are *i.i.d* $+1/-1$ *independently from their indices. If the solution of* (1.2) *is exact with high probability, then it is also exact with at least the same probability for the model in which the signs of entries in* $f^{(0)}$ *are fixed and* $\wedge(s_f)$ *are sampled from the Bernoulli model with parameter* $\rho$.

**Proof:**

This follows from a de-randomize technique introduced in [14] and the elimination theorem—theorem 2.1.1.∎

Finally, for the convenience of our proof and without loss of generality, in the remaining of this paper, we always assume that $1 \in s_x$, so that each row of matrix $\tilde{A}^*(s_x^c, s_f^c)$ forms a Steinhause sequence—a property that we need in section 2.5.

## 2.2 Dual certification

In this section, we'll prove theorem 1.1 base on the dual certification. The starting point of the proof is the so called "dual certification" as presented in the following lemma 2.2.1:

**Lemma 2.2.1 (dual certification [15])** $x^{(0)}$ *and* $f^{(0)}$ *is the unique solution of* (1.2) *if and only if the following condition holds:*

*There exists a vector* $h \in C^m$ *such that,*

$$\begin{cases} \lambda \tilde{A}^*(s_x, [m])h = \sigma_x, h(s_f) = \sigma_f \\ \left\| \lambda \tilde{A}^*(s_x^c, [m])h \right\|_\infty < 1, \left\| h(s_f^c) \right\|_\infty < 1 \end{cases} \quad (2.2.1)$$

*And matrix* $B = \left[ \lambda \tilde{A}([m], s_x), I_m([m], s_f) \right]$ *is full rank*

*Where* $\tilde{A}^*$ *denotes the conjugate matrix of* $\tilde{A}$.

**Proof.** this follows from a direct application of theorem 4 in [15]. ∎

According to [13], we have the below lemma 2.2.2.

**Lemma 2.2.2 ([13])** suppose there exists a $|s_x^c| + |s_f^c|$ dimensional vector q which satisfies,

$$\begin{bmatrix} \lambda \tilde{A}^*(s_x^c, s_f^c) & I_{|s_x^c|} \\ \lambda \tilde{A}^*(s_x, s_f^c) & 0 \end{bmatrix} q = \begin{bmatrix} -\lambda \tilde{A}^*(s_x^c, s_f)\sigma_f \\ \sigma_x - \lambda \tilde{A}^*(s_x, s_f)\sigma_f \end{bmatrix} \quad (2.2.2)$$

*And* $\|q\|_\infty < 1$, *where* $I_{|s_x^c|}$ *denotes a* $|s_x^c| \times |s_x^c|$ *identity matrix. Then there exists a m-dimensional vector h as defined in* (2.2.3) *satisfies* (2.2.1).

$$h(s_f) = \sigma_f, h(s_f^c) = q([|s_f^c|]) \tag{2.2.3}$$

**Proof.** According to (2.2.3) and (2.2.2), we have,

$$\begin{cases} \lambda \tilde{A}^*\left(s_x^c, s_f^c\right) h\left(s_f^c\right) + q\left(\left|s_f^c\right|+1:\left|s_f^c\right|+\left|s_x^c\right|\right) = -\lambda \tilde{A}^*\left(s_x^c, s_f\right) h\left(s_f\right) \\ \lambda \tilde{A}^*\left(s_x, s_f^c\right) h\left(s_f^c\right) = \sigma_x - \lambda \tilde{A}^*\left(s_x, s_f\right) h\left(s_f\right) \end{cases}$$

Which implies that:

$$\begin{cases} q\left(\left|s_f^c\right|+1:\left|s_f^c\right|+\left|s_x^c\right|\right) = -\lambda \tilde{A}^*\left(s_x^c, [m]\right) h \\ \lambda \tilde{A}^*\left(s_x, [m]\right) h = \sigma_x \end{cases} \tag{2.2.4}$$

Combining (2.2.4), (2.2.2) and the fact that $\|q\|_\infty < 1$ proves the conclusion of the lemma. ∎

In summary, if we can prove the matrix B defined in lemma 2.2.1 is full rank and there exist a vector q as stated in lemma 2.2.2 with high probability, then we can reach a final proof of theorem 1.1 according to lemma 2.2.1. The proof of B in lemma 2.2.1 being full rank is simple and it is provided in appendix D, while the proof of the existence of q requires more efforts, we outline the proof steps of the existence of q in next section, with the necessary supporting lemmas provided in appendix A ~ appendix C.

## 2.3 Existence of q in lemma 2.2.2

We achieve the goal of this section in 2 phases: in the first phase, we obtain a $q_0$ which approximately satisfies (2.2.2), as described below in (2.3.1).

For notational convenience, we denote the matrix $\begin{bmatrix} \lambda \tilde{A}^*\left(s_x^c, s_f^c\right) & I_{|s_x^c|} \\ \lambda \tilde{A}^*\left(s_x, s_f^c\right) & 0 \end{bmatrix}$ in (2.2.2) of lemma 2.2.2 as $\Phi$, and denote the vector $\begin{bmatrix} -\lambda \tilde{A}^*\left(s_x^c, s_f\right) \sigma_f \\ \sigma_x - \lambda \tilde{A}^*\left(s_x, s_f\right) \sigma_f \end{bmatrix}$ in (2.2.2) as $w$.

$$\begin{cases} \Phi q_0 = w, \; \|q_0([|s_1|])\|_2 \leq C_1 \sqrt{|s_x|} \sqrt{\ln\left(\dfrac{2n}{\varepsilon}\right)} \\ q_0\left(|s_1|+[|s_2|]\right) = 0, \; \|q_0\left(|s_f^c|+[|s_x^c|]\right)\|_\infty < 1/2 \end{cases} \tag{2.3.1}$$

Where $s_1, s_2$ are subsets of $s_f^c$ which satisfy, $s_1 \cup s_2 = s_f^c$, $|s_1| = |s_2|$, and $\wedge(s_1), \wedge(s_2)$

are random subset of $[n]$. $C_1$ is some positive constant to be determined later. The $q_0$ in (2.3.1) is called "approximately feasible" because $\|q_0\|_\infty$ can be larger than one (which violates conditions required in lemma 2.2.2) since $\|q_0([|s_1|])\|_\infty > 1$ can happen. Such $q_0$ which satisfies (2.3.1) is obtained by a golfing scheme as is described in section 2.4.

In the second phase, we prove the existence of a $\Delta q$, such that:

$$\Phi(q_0 + \Delta q) = w, \|q_0 + \Delta q\|_\infty < 1 \qquad (2.3.2)$$

And such $\Delta q$ is detailed in section 2.5.

For convenience, we state the results in section 2.4 and 2.5 formally as theorem 2.4.1 and theorem 2.5.1 as in section 2.4, section 2.5, respectively. Putting these theorems together with lemma 2.2.2 and lemma 2.2.1, it naturally leads to the conclusion of theorem 1.1.

## 2.4 Constructing $q_0$ satisfies (2.3.1) through a golfing scheme

Let $A_{|\lambda} = [\lambda \tilde{A}, I_m]$, then we propose the below golfing scheme to construct $q_0$. Firstly, we partition $s_f^c$ into 2 subset with equal sizes as mentioned in previous section: $s_f^c = s_1 \cup s_2$, $|s_1| = |s_2|$ so that $\wedge(s_1)$ and $\wedge(s_2)$ are both random subset of $[n]$.

**Golfing scheme 2.1 (constructing $q_0$).**

0. "initialization": let $h = 0 \in C^m$.

1. "Hitting $\sigma_f$": Construct a $\Delta h^{(0)}$, such that $\Delta h^{(0)}(s_f) = \sigma_f, \Delta h^{(0)}(s_f^c) = 0$, set $h = h + \Delta h^{(0)}$, $\Delta u^{(0)} = A_{|\lambda}^* \Delta h^{(0)}$, $u^{(0)} = A_{|\lambda}^* h$ and $w^{(0)} = \sigma_x - u^{(0)}(s_x)$.

2. "Approaching $\sigma_x$": for $j = 1, \ldots L$, construct a vector $\Delta h^{(j)}$, such that

$$\Delta h^{(j)}(\wedge_j) = \frac{m}{\lambda^2 |\wedge_j|} A_{|\lambda}(\wedge_j, s_x) w^{(j-1)} \quad , \quad \Delta h^{(j)}(\wedge_j^c) = 0 \quad \text{,let} \quad h = h + \Delta h^{(j)} \quad ,$$

$$\Delta u^{(j)} = A_{|\lambda}^* \Delta h^{(j)}, \quad u^{(j)} = A_{|\lambda}^* h \quad \text{and} \quad w^{(j)} = w^{(j-1)} - u^{(j)}(s_x).$$

3. "Hitting $\sigma_x$": construct a vector $\Delta h^{(L+1)}$, such that $\Delta h^{(L+1)}(s_1) = A_{|\lambda}^\dagger(s_1, s_x) w^{(L)}$, $\Delta h^{(L+1)}(s_1^c) = 0$, let $h = h + \Delta h^{(L+1)}$, $\Delta u^{(L+1)} = A_{|\lambda}^* \Delta h^{(L+1)}$ and $u = A_{|\lambda}^* h$, where $A_{|\lambda}^\dagger(s_1, s_x) = A_{|\lambda}(s_1, s_x)\left(A_{|\lambda}^*(s_x, s_1) A_{|\lambda}(s_1, s_x)\right)^{-1}$ denotes the Penron-Moore inverse of matrix $A_{|\lambda}(s_1, s_x)$.

4. $q_0\left(|s_f^c| + [|s_x^c|]\right) = \lambda \tilde{A}^*(s_x^c, [m]) h = u(s_x^c)$, $q_0\left([|s_1|]\right) = h(s_1)$, $q_0\left(|s_1| + [|s_2|]\right) = 0$.

In the 2$^{nd}$ step of golfing scheme above, we have $\bigcup_{j=1}^{L} \wedge_j = s_1$, with all $\wedge(\wedge_j)$ are disjoint random subset of $[n]$, for notational convenience, we denote $m_j = |\wedge_j|, j = 1, \ldots, L$.

To simplify the proof and our discussion in this paper, without loose of generality, we set

$$m_1 = m_2 = c_g |s_x| \ln(4n) \ln(2\varepsilon^{-1}), m_i = c_g |s_x| \ln(2L\varepsilon^{-1}), i \geq 3 \qquad (2.4.1)$$

With $\ln(2L\varepsilon^{-1}) L \leq \ln(4n) \ln(2\varepsilon^{-1})$ [III], where $L = \lceil \ln(|s_x|)/2 \rceil + 2$, $c_g \geq 842$ is a positive constant to be decided later in lemma B.3. For simplicity and without loosing generality, we also assume $\varepsilon^{-1} = n$, in this way, if n sufficiently large, then the probability bounds appear on the lemmas and theorems of this paper which dependent on $\varepsilon$ would become arbitrary close to 1.

Under the assumptions as described above, we have, $|s_1| = \frac{1}{2}|s_f| \leq 3c_g |s_x| \ln(4n) \ln(2\varepsilon^{-1})$, on the other hand, we have $|s_1| \geq m_1 + m_2 = 2c_g |s_x| \ln(4n) \ln(2\varepsilon^{-1})$. Hence it follows that $|s_f| = \alpha c_g |s_x| \ln(4n) \ln(2\varepsilon^{-1})$ where $\alpha$ is a constant lies within the interval $(4, 6)$. Furthermore, by our assumption that the number of corruptions: $|s_f|$ can at most occupy a positive fraction of the total measurement m, denote this corruption fraction as $\gamma_c \in (0, 1)$ as a constant, we then have

---

[III] This is achievable, notice that $\ln(2L\varepsilon^{-1}) L = L\ln(2L) + L\ln(\varepsilon^{-1})$, with both terms $L\ln(2L)$ and $L\ln(\varepsilon^{-1})$ asymptotically smaller than $\ln(4n)\ln(2\varepsilon^{-1})$, under the assumption that $\varepsilon^{-1} = n$.

$$m = c_m |s_x| \ln(4n) \ln(2\varepsilon^{-1}) \qquad (2.4.2)$$

With $c_m = \dfrac{\alpha}{1-\gamma_c} c_g$ to be some constant where $\gamma_c \in (0,1)$ and $\alpha \in (4,6)$.

We summarize our conclusion of this section in the below theorem 2.4.1.

**Theorem 2.4.1** *If* $m = c_m |s_x| \ln(4n) \ln(2\varepsilon^{-1})$, $\ln(4n) \geq \dfrac{\sqrt{3}}{2}\sqrt{\dfrac{1}{1-\gamma_c}}$, *setting*

$\lambda = \dfrac{c_\lambda}{\sqrt{\ln(2n/\varepsilon)}}$ *In (1.2), with positive constant* $c_\lambda \leq \dfrac{\sqrt{2}}{16}$, *then after the golfing scheme 2.1,*

*we have a* $q_0$ *which satisfies* (2.3.1)

$$\begin{cases} \Phi q_0 = w, \; \left\| q_0\left([|s_1|]\right) \right\|_2 \leq 2.1\beta \sqrt{|s_x|} \sqrt{\ln\left(\dfrac{2n}{\varepsilon}\right)} \\ q_0\left(|s_1| + [|s_2|]\right) = 0, \; \left\| q_0\left(|s_f^c| + [|s_x^c|]\right) \right\|_\infty < 1/2 \end{cases}$$

*With probability at least* $1-6\varepsilon$. *Where* $c_m = \dfrac{\alpha}{1-\gamma_c} c_g$, *with* $c_g \geq 842$, $\alpha \in (4,6)$ *to be*

*constants,* $\gamma_c \in (0,1)$ *is a constant denoting the corruption fraction,* $\beta = \dfrac{1}{10 c_\lambda} \sqrt{\dfrac{3\alpha}{1-\gamma_c}} \sqrt{\dfrac{3}{2}\dfrac{9}{8}}$

*is a positive constant.*

**Proof:**

Firstly, $\Phi q_0 = w$, $q_0\left(|s_1| + [|s_2|]\right) = 0$ holds according to the golfing scheme, secondly,

$\left\| q_0\left([|s_1|]\right) \right\|_2 \leq 2.1\beta |s_x| \sqrt{\ln\left(\dfrac{2n}{\varepsilon}\right)}$ holds with high probability according to theorem C.1,

finally, according to theorem B.1, we have,

$$\left\| q_0\left(|s_f^c| + [|s_x^c|]\right) \right\|_\infty = \left\| u(s_x^c) \right\|_\infty < 1/2$$

Holds also with high probability, by applying a union bound we prove the theorem. ∎

## 2.5 proof of the existence of $\Delta q$ in (2.3.2)

Given a vector $q_0$ satisfies (2.3.1) as constructed in the previous section 2.4 through golfing scheme, our goal in this section is therefore to prove the existence of a "modification" vector

$\Delta q$ satisfies (2.3.2), so that $q_0 + \Delta q$ satisfies the dual certification as elaborated in lemma 2.2.2, and in this way, the existence of a dual vector that satisfies the dual certificate conditions in lemma 2.2.2 is verified. The main result of this section is stated formally as in theorem 2.5.1 as below,

**Theorem 2.5.1** If $\sqrt{\dfrac{|s_f^c|}{2}} - |s_x| > 4.2\sqrt{3}\beta\sqrt{|s_x|}\sqrt{\ln\left(\dfrac{2n}{\varepsilon}\right)}$ and $m \geq 318 c_\lambda^2 \beta^2 \ln(2n\varepsilon^{-1})|s_x|$, then with probability at least $1 - 6\varepsilon$, there exists a $\Delta q$, such that $\Phi \Delta q = 0$, with $\Delta q([|s_1|]) = -q_0([|s_1|])$, $\|\Delta q(|s_1| + [|s_2|])\|_\infty < 1/2$, $\|\Delta q(|s_f^c| + [|s_x^c|])\|_\infty < 1/2$. Where $q_0$ is the vector constructed by the golfing scheme 2.1.

We decompose the proof of theorem 2.5.1 into 2 steps, in the first step in lemma 2.5.3, we devote to prove that there exists a $|s_2|$ dimensional vector $y$, with $\|y\|_\infty < 1/2$, which satisfies,

$$-\tilde{A}^*(s_x, s_1) q_0([|s_1|]) + \tilde{A}^*(s_x, s_2) y = 0 \tag{2.5.1}$$

After that, we let $\Delta q(|s_1| + [|s_2|]) = y$, and then in the second step in lemma 2.5.4, we dedicate to find a $|s_x^c|$ dimensional vector $v$ with $\|v\|_\infty < 1/2$, satisfying,

$$-\lambda \tilde{A}^*(s_x^c, s_1) q_0([|s_1|]) + \lambda \tilde{A}^*(s_x^c, s_2) y + v = 0 \tag{2.5.2}$$

Then we let $\Delta q(|s_f^c| + [|s_x^c|]) = v$.

After the above 2 phases, a valid dual vector $q = q_0 + \Delta q$ that satisfies the dual certification (see lemma 2.2.2) is therefore established, to see this, firstly $\Phi \Delta q = 0$ is easy to verify according to (2.5.1) and (2.5.2), let $q = q_0 + \Delta q$, then obviously we have $\Phi q = w$, furthermore, we have

$$\|q([|s_1|])\|_\infty = 0 < 1,$$

$$\|q(|s_1| + [|s_2|])\|_\infty \leq \|q_0(|s_1| + [|s_2|])\|_\infty + \|\Delta q(|s_1| + [|s_2|])\|_\infty < 0 + \dfrac{1}{2} < 1, \text{ and}$$

$$\|q(|s_f^c| + [|s_x^c|])\|_\infty \leq \|q_0(|s_f^c| + [|s_x^c|])\|_\infty + \|\Delta q(|s_f^c| + [|s_x^c|])\|_\infty < \dfrac{1}{2} + \dfrac{1}{2} = 1, \text{ which implies}$$

$\|q\|_\infty < 1$, therefore, the dual certification conditions in lemma 2.2.2 is verified on $q$.

**Lemma 2.5.1** (subgradient of $\|\cdot\|_\infty$) let $g_0 := \nabla_{x=x_0} \|x\|_\infty$ denotes the sub-gradient of the convex function $\|\cdot\|_\infty$ at $x = x_0$ where $x$ denotes a n-dimension vector variable, $x_0 \in R^{+n}$ denotes a constant vector whose entries are non-negative. Then $g_0(i) = 1$ if and only if $x_0(i) = \|x_0\|_\infty$.

**Proof:**

This follows by the convexity of function $\|\cdot\|_\infty$, in which we have, $\|x_0 + \Delta x\|_\infty \geq \|x_0\|_\infty + g_0^T \Delta x$ holds for any $\Delta x \in R^n$. ∎

**Lemma 2.5.2** let $y$ be a vector that satisfies (2.5.1), then with probability at least $1 - 5\varepsilon$, we have $\|y\|_2 \leq 2.1\sqrt{3}\beta\sqrt{|s_x|}\sqrt{\ln\left(\frac{2n}{\varepsilon}\right)}$.

**Proof:**

Firstly, notice that $\|q_0([|s_1|])\|_2 = \|h(s_f^c)\|_2$, according to lemma A.2 and theorem C.1, the follow events,

$$\|\tilde{A}^*(s_x, s_1)\|_{2\to 2} \leq \sqrt{\frac{|s_1|}{m}}\sqrt{\frac{3}{2}}, \|\tilde{A}^*(s_x, s_2)\|_{2\to 2} \geq \sqrt{\frac{|s_2|}{m}}\sqrt{\frac{1}{2}}, \text{ and}$$
$$\|q_0([|s_1|])\|_2 \leq 2.1\beta\sqrt{|s_x|}\sqrt{\ln\left(\frac{2n}{\varepsilon}\right)}$$
(2.5.3)

Hold with probability at least $1 - 5\varepsilon$, let $z$ denotes $\tilde{A}^*(s_x, s_1)q_0([|s_1|])$, $M$ denotes $\tilde{A}^*(s_x, s_2)$ we therefore have,

$$z = My \tag{2.5.4}$$

now we bound the probability of the inverse event, that is,

$$\|y\|_2 > 2.1\sqrt{3}\beta\sqrt{|s_x|}\sqrt{\ln\left(\frac{2n}{\varepsilon}\right)} \tag{2.5.5}$$

If event (2.5.5) holds, according to (2.5.3), it implies,

$$\|z\|_2 = \|My\|_2 > \sqrt{\frac{|s_2|}{m}}\sqrt{\frac{3}{2}} 2.1\beta\sqrt{|s_x|}\sqrt{\ln\left(\frac{2n}{\varepsilon}\right)} \tag{2.5.6}$$

But by (2.5.3), with probability at least $1-5\varepsilon$, it has,

$$\|z\|_2 \leq \sqrt{\frac{|s_2|}{m}}\sqrt{\frac{3}{2}}2.1\beta\sqrt{|s_x|}\sqrt{\ln\left(\frac{2n}{\varepsilon}\right)}$$

Which leads to a contradiction with probability at least $1-5\varepsilon$, hence the conclusion of this lemma follows.∎

**Lemma 2.5.3** If $\sqrt{\frac{|s_f^c|}{2}}-|s_x| > 4.2\sqrt{3}\beta\sqrt{|s_x|}\sqrt{\ln\left(\frac{2n}{\varepsilon}\right)}$, then with probability at least $1-5\varepsilon$, there exists a $|s_2|$ dimensional vector $y \in C^n$, which satisfies (2.5.1). With $\|y\|_\infty < \frac{1}{2}$. Here the $\ell_\infty$ norm of a complex vector means the maximum value of the magnitudes of entries in the vector, e.g., $\|y\|_\infty = \max\{|y(i)|, i \in [n]\}$.

**Proof:**

We argue by contradiction, for notational convenience, as in the previous lemma, let $z$ denotes the vector $\tilde{A}^*(s_x, s_1)q_0(\lceil |s_1|\rceil)$, $M$ denotes the matrix $\tilde{A}^*(s_x, s_2)$, then (2.5.1) is simplified as,

$$z - My = 0 \tag{2.5.7}$$

Suppose the inverse of the conclusion holds, e.g., for every $y$ satisfying (2.5.7), we have $\|y\|_\infty > 1/2$, then it implies that $\|y^*\|_\infty > 1/2$, where $y^*$ is the solution of the following optimization:

$$\text{minimize}_y \|y\|_\infty, \text{ s.t. } z - My = 0 \tag{2.5.8}$$

For the convenience of the proof, we reformulate (2.5.8) as

$$\text{minimize}_y \|y\|_\infty, \text{ s.t. } z - \tilde{M}y = 0 \tag{2.5.9}$$

Where the $i^{th}$ column of $\tilde{M}$ is formed by multiplying the $i^{th}$ column of $M$ by a complex value $\text{sgn}(y^*(i))$ [IV] for all $i \in [n]$. In this manner, since $y^*$ is an optimal solution of (2.5.8), one can conclude that a non-negative vector $\tilde{y}^*$ with $\tilde{y}^*(i) = |y^*(i)|, i \in [n]$ is a solution of (2.5.9).

By the KKT condition, there exist a $|s_x|$ dimensional dual variables vector $\mu$, such that,

---

[IV] For a complex value $z = re^{i\theta}$ with $r > 0$, we denote $|z| = r$ as its magnitude and $\text{sgn}(z) = e^{i\theta}$ as its sign.

$$\nabla_{y=\tilde{y}^*} \|y\|_\infty - \tilde{M}^* \mu = 0 \qquad (2.5.10)$$

Let $g := \nabla_{y=\tilde{y}^*} \|y\|_\infty$ denotes the subgradient of $\|y\|_\infty$ at $\tilde{y}^*$, then we rewrite (2.5.10) as,

$$g - \tilde{A}\left(|s_1| + [|s_2|], s_x\right)\tilde{\mu} = 0 \qquad (2.5.11)$$

Where in (2.5.11), $\tilde{\mu}(i) = \mu(i)\operatorname{sgn}(y^*(i)), i \in [n]$, by lemma A.5 [16], (2.5.11) implies that,

$$\|g\|_0 + n - |s_f^c|/2 + |s_x| \geq 1 + n \qquad (2.5.12)$$

Which gives,

$$\|g\|_0 \geq \frac{|s_f^c|}{2} - |s_x| \qquad (2.5.13)$$

According to the property of sub-gradient of $\|\cdot\|_\infty$ in lemma 2.5.1, (2.5.13) further implies that,

$$\|y^*\|_2 \geq \frac{1}{2}\sqrt{\frac{|s_f^c|}{2} - |s_x|} \qquad (2.5.14)$$

Since $\|y\|_2 \leq 2.1\sqrt{3}\beta\sqrt{|s_x|}\sqrt{\ln\left(\frac{2n}{\varepsilon}\right)}$ holds with probability at least $1-5\varepsilon$ according to lemma 2.5.2. Then if $\sqrt{\frac{|s_f^c|}{2} - |s_x|} > 4.2\sqrt{3}\beta\sqrt{|s_x|}\sqrt{\ln\left(\frac{2n}{\varepsilon}\right)}$, we reach at a contradiction with

probability at least $1-5\varepsilon$, hence the conclusion of this lemma holds. ∎

**Lemma 2.5.4** If $m \geq 318c_\lambda^2\beta^2 \ln\left(2n\varepsilon^{-1}\right)|s_x|$, then with probability at least $1-6\varepsilon$, there exists a vector $v$ which satisfies (2.5.2), with $\|v\|_\infty < 1/2$.

**Proof:**
Rewrite (2.5.2) as,

$$v = \lambda\tilde{A}^*\left(s_x^c, s_f^c\right)\left[-q_0\left([|s_1|]\right)^T, y^T\right]^T \qquad (2.5.15)$$

since by triangle rule, and lemma 2.5.2, with probability at least $1-5\varepsilon$, it holds,

$$\left\|\left[-q_0\left([|s_1|]\right)^T, y^T\right]^T\right\|_2 \leq \left\|q_0\left([|s_1|]\right)\right\|_2 + \|y\|_2 \leq 6.3\beta\sqrt{|s_x|}\sqrt{\ln\left(\frac{2n}{\varepsilon}\right)} \qquad (2.5.16)$$

Let vector $a$ denotes $\frac{\lambda}{\sqrt{m}}\left[-q_0\left([|s_1|]\right)^T, y^T\right]^T$, then, $v(i)$ can be interpreted as sum of a

Steinhaus' sequence[v] weighted by $a$, we have,

$$prob\left(|v(i)| \geq \frac{1}{2}\right) = prob\left(\left|\sum_{j=1}^{|s_f^c|} \epsilon_j a(j)\right| \geq \frac{1}{2}\right), i = 1, \ldots, |s_x^c| \quad (2.5.17)$$

Where $\epsilon_j, j = 1, \ldots, |s_f^c|$ in (2.5.17) interpreted as a Steinhaus' sequence.

According to lemma A.3, and notice that $\|a\|_2 \leq \frac{c_\lambda}{\sqrt{m}} 6.3\beta\sqrt{|s_x|}$, we have,

$$prob\left(|v(i)| \geq \frac{1}{2}\right) =$$
$$prob\left(\left|\sum_{j=1}^{|s_f^c|} \epsilon_j a(j)\right| \geq \frac{1}{2}\right) \leq 2\exp\left(-\frac{1}{318 c_\lambda^2 \beta^2} \frac{m}{|s_x|}\right) \leq \varepsilon/n \quad (2.5.18)$$

then conclusion of this lemma follows by applying a union bound on (2.5.18). ∎

**Proof of theorem 2.5.1:**
It follows by lemma 2.5.3~lemma 2.5.4 and then applying a union bound. ∎

**Proof of theorem 1.1:**
Following the conclusion of theorem 2.4.1 and theorem 2.4.2, we've successfully proven the existence of a dual vector $q$ that satisfies the dual certificate condition as elaborated in lemma 2.2, furthermore, according to theorem D.1, the matrix $\left[\lambda\tilde{A}([m], s_x), I_m([m], s_f)\right]$ is full rank, hence by lemma 2.1, the conclusion of the theorem holds by applying a union bound. ∎

# Appendix A preliminary lemmas

***Lemma A.1 (theorem 12.12 of [6])*** *Let $A \in C^{m \times n}$ be random sampling matrix associated to a Bounded Orthogonal System (BOS) with constant $K \geq 1$. Let $s \subset [n]$ be an indices set. Then, for $\delta \in (0,1)$, the normalized matrix $\tilde{A} = \frac{1}{\sqrt{m}} A$ satisfies $\left\|\tilde{A}^*([m], s)\tilde{A}([m], s) - I_{|s|}\right\|_{2 \to 2} \leq \delta$, with probability at least $1 - 2|s|\exp\left(-\frac{3m\delta^2}{8K^2|s|}\right)$.*

***Lemma A.2*** *if $\wedge$ and $\wedge(s)$ are random subsets of $[n]$, where $s$ is a subset of $[n]$ then,*

---
[v] Recalling that $1 \notin s_x^c$ by our assumption at the end of section 2.1.

$$\text{prob}\begin{pmatrix} \|\tilde{A}^*(s_x,s)\|_{2\to 2} \leq \sqrt{\frac{|s|}{m}}\sqrt{\frac{3}{2}}, \\ \|\tilde{A}^\dagger(s,s_x)\|_{2\to 2} \leq \sqrt{\frac{m}{|s|}}\sqrt{6} \end{pmatrix} \geq 1-2\varepsilon \quad (A.1)$$

holds for any $\varepsilon \in \left(0, \frac{1}{2}\right)$, provided that $|s| \geq \frac{32}{3}|s_x|\ln\left(\frac{2|s_x|}{\varepsilon}\right)$. Where $\wedge$ denotes the row indices of matrix $A$ that corresponding to the Fourier basis matrix, and $m = |\wedge|$

**Proof:**

Substituting the constant $K=1$ and letting $\delta = 1/2$ into lemma A.1, one has,

$$\text{prob}\left(\left\|\frac{m}{|s|}\tilde{A}^*(s_x,s)\tilde{A}(s,s_x) - I_{|s_x|}\right\|_{2\to 2} \leq \frac{1}{2}\right) \geq 1-2\varepsilon \quad (A.2)$$

provided that,

$$|s| \geq \frac{32}{3}|s_x|\ln\left(\frac{2|s_x|}{\varepsilon}\right) \quad (A.3)$$

(A.2) also implies the events,

$\|\tilde{A}^*(s_x,s)\|_{2\to 2} \leq \sqrt{\frac{|s|}{m}}\sqrt{\frac{3}{2}}$ and $\|\tilde{A}^\dagger(s,s_x)\|_{2\to 2} \leq \sqrt{\frac{m}{|s|}}\sqrt{6}$, which proves the conclusion of the lemma.∎

**Lemma A.3** (Hoeffding-type inequality for Steinhaus sums, corollary 8.10 in [6]) Let $a \in C^M$ and $\epsilon = (\epsilon_1,\ldots,\epsilon_M)$ be a Steinhaus sequence. For any $0 < \gamma < 1$,

$$\text{prob}\left(\left|\sum_{j=1}^M \epsilon_j a(j)\right| \geq u\|a\|_2\right) \leq \frac{1}{1-\gamma}\exp(-\gamma u^2) \text{ for all } u > 0.$$

**Lemma A.4** (Hoeffding-type inequality for Rademacher sum, corollary 8.8 in [6]) let $a \in C^M$ and $\epsilon = (\epsilon_1,\ldots,\epsilon_M)$ be a Rademacher sequence. Then, for u>0,

$$\text{prob}\left(\left|\sum_{j=1}^M \epsilon_j a(j)\right| \geq u\|a\|_2\right) \leq 2\exp\left(-\frac{u^2}{2}\right)$$

**Lemma A.5** ([16]) let $f$ be a n-dimensional non-zero complex vector, $\overset{\vee}{f}$ be the Fourier transform of $f$, suppose n is prime, then,

$$\|f\|_0 + \|\overset{\vee}{f}\|_0 \geq n+1$$

# Appendix B bounding $\|w^{(j)}\|_2$ and $\|u^{(j)}(s_x^c)\|_\infty$

**Lemma B.1** we have $prob\left\{\|u^{(0)}([n])\|_\infty < \frac{1}{8}\right\} \geq 1-\varepsilon$, provided that $\lambda = \frac{c_\lambda}{\sqrt{\ln(2n/\varepsilon)}}$, with positive constant $c_\lambda \leq \frac{\sqrt{2}}{16}$.

**Proof:**

By the definition of $u^{(0)}$ in the above golfing scheme, we have $u^{(0)}([n]) = \lambda \tilde{A}^*([n], s_f) \Delta h^{(0)}(s_f) = \lambda \tilde{A}^*([n], s_f) \sigma_f$, since the signs of $\sigma_f$ are within {+1/-1} uniformly and randomly, we can treat that $u^{(0)}(i), i \in [n]$ to be sum of a Randemacher's sequence or Steinhause's sequence, according to lemma A.3~A.4, with constant $\gamma = 1/2$, we have,

$$prob\left\{u^{(0)}(i) < \lambda\sqrt{\frac{|s_f|}{m}}\mu\right\} \geq 1 - 2\exp\left(-\frac{\mu^2}{2}\right) \quad (B.1)$$

Since $\sqrt{\frac{|s_f|}{m}} < 1$, letting $\mu = \frac{1}{8\lambda}$, (B.1) implies that,

$$prob\left\{u^{(0)}(i) < \frac{1}{8}\right\} \geq 1 - 2\exp\left(-\frac{1}{128\lambda^2}\right) \geq 1 - 2\exp\left(-\frac{\ln(2n/\varepsilon)}{128c_\lambda^2}\right) \quad (B.2)$$

applying a union bound to (B.2) yields the conclusion of the lemma. ∎

**Lemma B.2** $\|w^{(0)}\|_2 \leq \frac{9}{8}\sqrt{|s_x|}$, $\|u^{(0)}(s_x^c)\|_\infty < \frac{1}{8}$, with probability at least $1-\varepsilon$, provided the conditions of lemma B.1 holds.

**Proof:**
This follows from lemma B.1. ∎

**Lemma B.3** Under the setting of (2.4.1) and (2.4.2), let $c_\vartheta \geq 842$, then with probability at least $1-\frac{5}{2}\varepsilon$, one has $\left\|w^{(L)}\right\|_2 \leq \frac{1}{12}\frac{1}{\ln(4n)}$, and $\left\|u^{(L)}\left(s_x^c\right)\right\|_\infty < \frac{1}{4}$.

**Proof:**

We follow the notions and arguments of [6] p.399, there are only 2 differences here: the first is that we replace the constant "e" (which is used to calculate $r_n$ and $t_n$, $1 \leq n \leq L$) in [6] p.399 with a constant $\vartheta$ that to be determined; the second difference is that $u^{(0)}\left(s_x^c\right)$ here is not zero vector but whose $\ell_\infty$ norm is bounded by 1/8. Let $\vartheta = 10$ and

$$c_\vartheta = 8\vartheta^2\left(1+\left(\frac{1}{\sqrt{8}}+\frac{1}{6}\right)\Big/\vartheta\right) \approx 842 \quad , \quad \text{with} \quad \varepsilon \in (0, 1/6) \quad \text{setting parameters}$$

$r_1 = r_2 = \frac{1}{2\vartheta\sqrt{\ln(4n)}}, r_i = \frac{1}{2\vartheta}, i = 3, \ldots, L$, $t_1 = t_2 = \frac{1}{\vartheta\sqrt{|s_x|}}, t_i = \frac{\ln(4n)}{\vartheta\sqrt{|s_x|}}, i \geq 3$, which are chosen dependent on $\vartheta$ similar to those of [1 p.399].

Then we have, $r_1' = r_2' \leq 1/\left(\vartheta\sqrt{\ln(4n)}\right)$ and $r_i' < \vartheta^{(-1)}, i = 3, \cdots, L$. Where $r_i' = \sqrt{\frac{|s_x|}{m_i}} + r_i$.

The meaning of above notations $r_i$, $r_i'$ and $t_i$, for $1 \leq i \leq L$ follow exactly from [6] p.399. Following the reasoning of [6] p.399, we have,

$$\left\|w^{(L)}\right\|_2 \leq \left\|w^{(0)}\right\|_2 \prod_{i=1}^{L} r_i' \leq \frac{9}{8}\sqrt{|s_x|}\vartheta^{-\frac{\ln(|s_x|)}{2}-2}\frac{1}{\ln(4n)} < \frac{9}{8}\vartheta^{-2} < \frac{1}{12\ln(4n)} \quad (B.3)$$

$$\left\|u^{(L)}\left(s_x^c\right)\right\|_\infty \leq$$

$$\left\|u^{(0)}\left(s_x^c\right)\right\|_\infty + \frac{9}{8}\vartheta^{-1}\left(1 + \frac{1}{\vartheta\sqrt{\ln(4n)}} + \sum_{i=2}^{L-1}\vartheta^{-i}\right) \leq \frac{1}{8} + \frac{9}{8}\frac{\vartheta^{-1}}{1-\vartheta^{-1}} < \frac{1}{4} \quad (B.4)$$

With (B.3) holds with probability at least $1-\frac{\varepsilon}{2}$, and (B.4) holds with probability at least $1-\varepsilon$, together with the result of lemma B.2 and then applying a union bound gives the conclusion of the lemma. ∎

**Lemma B.4** $\left\|\Delta u^{(L+1)}\left(s_x^c\right)\right\|_\infty \leq \frac{1}{4}$ with probability at least $1-6\varepsilon$ provided that

$\ln(4n) \geq \frac{\sqrt{3}}{2}\sqrt{\frac{1}{1-\gamma_c}}$ and conditions in lemma B.3 holds, where $\gamma_c = \frac{|s_f|}{m} \in (0,1)$ denotes the corruption ratio.

**Proof:**

We bound $\left\|\Delta u^{(L+1)}(s_x^c)\right\|_\infty$ by its $\ell_2$ norm:

$$\left\|\Delta u^{(L+1)}(s_x^c)\right\|_\infty \leq \left\|\Delta u^{(L+1)}(s_x^c)\right\|_2 \leq$$
$$\left\|\tilde{A}_{|\lambda}^* \Delta h^{(L+1)}\right\|_2 \leq \left\|\tilde{A}_{|\lambda}^*(s_x^c, s_1)\right\|_{2\to2} \left\|\tilde{A}_{|\lambda}^\dagger(s_1, s_x)\right\|_{2\to2} \left\|w^{(L)}\right\|_2 \quad (B.5)$$

By lemma B.3, we have,

$$prob\left\{\left\|w^{(L)}\right\|_2 < \frac{1}{12}\frac{1}{\ln(4n)}\right\} \geq 1 - \frac{5}{2}\varepsilon \quad (B.6)$$

Furthermore,

$$\left\|\tilde{A}_{|\lambda}^*(s_x^c, s_1)\right\|_{2\to2} \leq \lambda \left\|\tilde{A}^*\right\|_{2\to2} = \lambda \quad (B.7)$$

According to lemma A.2, we further have,

$$prob\left\{\left\|\tilde{A}_{|\lambda}^*(s_x^c, s_1)\right\|_{2\to2} \left\|\tilde{A}_{|\lambda}^\dagger(s_1, s_x)\right\|_{2\to2} \leq \sqrt{\frac{12}{1-\gamma_c}}\right\} \geq 1 - 2\varepsilon \quad (B.8)$$

putting together (B.6) and (B.8) and then applying a union bound yields the desired conclusion of this lemma. ∎

**Theorem B.1.** Under the setting of (2.4.1) and (2.4.2) with $c_g \geq 842$, let $\lambda = \frac{c_\lambda}{\sqrt{\ln(2n/\varepsilon)}}$ in (1.2), with positive constant $c_\lambda \leq \frac{\sqrt{2}}{16}$, if $\ln(4n) \geq \frac{\sqrt{3}}{2}\sqrt{\frac{1}{1-\gamma_c}}$, then with probability at least $1-6\varepsilon$, $\left\|u(s_x^c)\right\|_\infty < 1/2$ holds. Where $\gamma_c = \frac{|s_f|}{m} \in (0,1)$ denotes the corruption ratio.

**Proof:**
This follows by lemma B.3 and lemma B.4 and then apply a union bound. ∎

# Appendix C. bounding $\left\|h(s_f^c)\right\|_2$

**Theroem C.1** If conditions in theorem B.1 holds, then $\left\|h(s_f^c)\right\|_2 < 2.1\beta\sqrt{|s_x|}\sqrt{\ln\left(\frac{2n}{\varepsilon}\right)}$ with

probability at least $1-5\varepsilon$, where $\beta = \dfrac{1}{10c_\lambda}\sqrt{\dfrac{3\alpha}{1-\gamma_c}}\sqrt{\dfrac{3}{2}}\dfrac{9}{8}$, is a positive constant.

**Proof:**

Firstly, according to the golfing scheme,

$$\left\|\Delta h^{(j)}\right\|_2 \leq \left\|\frac{m}{\lambda^2 m_j}\tilde{A}_{|\lambda}\left(\wedge_j, s_x\right)\right\|_{2\to 2}\left\|w^{(j-1)}\right\|_2, j=1,\ldots,L \tag{C.1}$$

According to the setting (2.4.1) and (2.4.2) in section 2.4, and by lemma A.2 we have,

$$\left\|\frac{m}{\lambda^2 m_j}\tilde{A}_{|\lambda}\left(\wedge_j, s_x\right)\right\|_{2\to 2} \leq \frac{1}{\lambda}\sqrt{\frac{m}{m_j}}\left\|\sqrt{\frac{m}{m_j}}\tilde{A}\left(\wedge_j, s_x\right)\right\|_{2\to 2} \leq$$

$$\begin{cases} \dfrac{1}{c_\lambda}\sqrt{\dfrac{3\alpha}{1-\gamma_c}}\sqrt{\ln\left(\dfrac{2n}{\varepsilon}\right)}\sqrt{3/2}, j=1,2 \\ \dfrac{1}{c_\lambda}\sqrt{\dfrac{3\alpha}{1-\gamma_c}}\sqrt{\ln\left(\dfrac{2n}{\varepsilon}\right)}\sqrt{3/2}\sqrt{\ln 4n}, j=3,\ldots L \end{cases} \tag{C.2}$$

holds with probability $1-2\varepsilon$.

Furthermore, by the arguments of lemma B.3,

$$\left\|w^{(0)}\right\|_2 \leq \frac{9}{8}\sqrt{|s_x|} \tag{C.3}$$

holds with probability at least $1-\varepsilon$, combining (C.1), (C.2) and (C.3) gives,

$$\left\|\Delta h^{(1)}\right\|_2 \leq \frac{1}{c_\lambda}\sqrt{\frac{3\alpha}{1-\gamma_c}}\sqrt{\ln\left(\frac{2n}{\varepsilon}\right)}\sqrt{\frac{3}{2}}\frac{9}{8}\sqrt{|s_x|} \leq \beta\sqrt{\ln\left(\frac{2n}{\varepsilon}\right)}\sqrt{|s_x|} \tag{C.4}$$

$$\left\|\Delta h^{(i)}\right\|_2 \leq \frac{1}{c_\lambda}\sqrt{\frac{3\alpha}{1-\gamma_c}}\sqrt{\ln\left(\frac{2n}{\varepsilon}\right)}\sqrt{3/2}\sqrt{\ln 4n}\left\|w^{(0)}\right\|_2 \prod_{j=1}^{i-1}r_i' \leq$$

$$\beta\sqrt{\ln\left(\frac{2n}{\varepsilon}\right)}\sqrt{|s_x|}\vartheta^{-(i-1)}, i=2,\ldots,L \tag{C.5}$$

Which holds with probability at least $1-3\varepsilon$ by applying a union bound. Where positive constant $\beta$ in (C.4) is defined the quantity $\dfrac{1}{9c_\lambda}\sqrt{\dfrac{3\alpha}{1-\gamma_c}}\sqrt{\dfrac{3}{2}}\dfrac{9}{8}$. Using similar argument as in above, we can bound $\left\|\Delta h^{(L+1)}\right\|_2$ as following,

$$\left\|\Delta h^{(L+1)}\right\|_2 = \left\|A_{|\lambda}^\dagger\left(s_1,s_x\right)w^{(L)}\right\|_2 \leq \left\|A_{|\lambda}^\dagger\left(s_1,s_x\right)\right\|_{2\to 2}\left\|w^{(L)}\right\|_2 \leq$$

$$\frac{1}{c_\lambda}\sqrt{\ln\left(\frac{2n}{\varepsilon}\right)}\sqrt{\frac{\alpha}{1-\gamma_c}}\sqrt{6}\frac{1}{12} \leq \beta\sqrt{\ln\left(\frac{2n}{\varepsilon}\right)} \tag{C.6}$$

Which holds with probability at least $1-3\varepsilon$. Where the second inequality of (C.5) follows from lemma A.3.

Choose $\vartheta = 10$ as in the previous section, and then applying the triangle rule, we bound $\|h(s_f^c)\|_2$ as following,

$$\|h(s_f^c)\|_2 \leq \|\Delta h^{(L+1)}\|_2 + \sum_{i=1}^{L} \|\Delta h^{(i)}\|_2$$
$$\leq \beta\sqrt{\ln\left(\frac{2n}{\varepsilon}\right)} + \beta\sqrt{|s_x|}\sqrt{\ln\left(\frac{2n}{\varepsilon}\right)}\left(\frac{1-\vartheta^{-L}}{1-\vartheta}\right) \quad (C.7)$$
$$\leq \beta\sqrt{\ln\left(\frac{2n}{\varepsilon}\right)} + 1.1\beta\sqrt{|s_x|}\sqrt{\ln\left(\frac{2n}{\varepsilon}\right)} < 2.1\beta\sqrt{|s_x|}\sqrt{\ln\left(\frac{2n}{\varepsilon}\right)}$$

Which holds with probability at least $1-5\varepsilon$, which proves the conclusion of this theorem. ∎

# Appendix D. Proving the full rank of $\left[\lambda\tilde{A}([m],s_x), I_m([m],s_f)\right]$

**Theorem D.1 (theorem A.3.1 of [17])** if $\wedge$ and $\wedge(s_f)$ are random subsets of $[n]$,

$$|s_f^c| \geq \frac{32}{3}|s_x|\ln\left(\frac{2|s_x|}{\varepsilon}\right) \quad , \quad \text{then with probability at least} \quad 1-\varepsilon \quad , \quad \text{matrix}$$

$B = \left[\lambda\tilde{A}([m],s_x), I_m([m],s_f)\right]$ is full rank.

**Proof:**

to show $B$ is full rank, it's sufficient to show that matrix $C = B^*B$ is non-singular, in other words, the minimal eigenvalue of $C$ is positive. Since,

$$C = \begin{bmatrix} \lambda^2\tilde{A}^*(s_x,[m])\tilde{A}([m],s_x) & \lambda\tilde{A}^*(s_x,s_f) \\ \lambda\tilde{A}(s_f,s_x) & I_{|s_f|} \end{bmatrix} \quad (D.1)$$

By Schur complement decomposition, one has,

$$HCH^T = \begin{bmatrix} \lambda^2\tilde{A}^*(s_x,s_f^c)\tilde{A}(s_f^c,s_x) & 0 \\ 0 & I_{|s_f|} \end{bmatrix} \quad (D.2)$$

Where $H = \begin{bmatrix} I_{|s_x|} & -\lambda\tilde{A}^*(s_x,s_f) \\ 0 & I_{|s_f|} \end{bmatrix}$ is a non-singular matrix, by lemma A.2.1, one has

$$prob\left(\|\lambda^2\tilde{A}^*(s_x,s_f^c)\tilde{A}(s_f^c,s_x)\|_{2\to 2} \geq \frac{\lambda^2}{2}\right) \geq 1-\varepsilon \quad (D.3)$$

Holds if $|s_f^c| \geq \frac{32}{3}|s_x|\ln\left(\frac{2|s_x|}{\varepsilon}\right)$.

The conclusion of this theorem is proved by combining (D.2) and (D.3). ∎